\documentclass[preprint,preprintnumbers,amsmath,amssymb]{revtex4}

\usepackage{graphicx}
\usepackage{dcolumn}
\usepackage{bm}

\begin{document}

\title{Effects of coating rate on morphology of copper surfaces}

\author{S. Motamen, M. Vahabi}
\address{Department of Physics, Shahid Beheshti University, G. C.,
Evin, Tehran 19839, Iran}

\author{G. R. Jafari\footnote{Corresponding author}}

\address{Department of Physics, Shahid Beheshti University,
G.C., Evin, Tehran 19839, Iran \\
School of Nano-Science, Institute for Research in Fundamental Sciences (IPM), P.O. Box 19395-5531, Tehran, Iran\\
g\_jafari@sbu.ac.ir}


\begin{abstract}

We have used standard fractal analysis and Markov approach to obtain
further insights on roughness and multifractality of different
surfaces. The effect of coating rates on generating topographic
rough surfaces in copper thin films with same thickness has been
studied using atomic force microscopy technique (AFM). Our results
show that by increasing the coating rates, correlation length (grain
sizes) and Markov length are decreased and roughness exponent is
decreased and our surfaces become more multifractal. Indeed, by
decreasing the coating rate, the relaxation time of embedding the
particles is increased.

\end{abstract}
\maketitle

\section{Introduction}

Surface, which is the first interface of a material, has an
important role in the interaction of matter with environment.
Physical and chemical properties of the surfaces are not only
determined by the material properties but also to a significant
degree by the topography. No surface is perfectly flat and every
surface depending on the scale it is observed, has a certain amount
of roughness. Surface roughness is one of the important properties
of the surfaces that affect many other properties of the surface
like adhesion, friction and contact, reflection or scattering
\cite{Paillet,Yaish,adhesion,adfric,friction,light,Mahdavi}. The
roughness effect appears in various devices such as field emission
devices \cite{Karabutov}, sensors \cite{sensor}, self-cleaning
materials \cite{selfclean}. It should be noted that roughness is not
an intrinsic property of the surface. Indeed, it depends on the
scale of observation. In other words, when we observe a surface from
different scales different roughness could be obtained.

In this article, we investigate the effect of coating rate on
statistical properties of our prepared surfaces. These surfaces have
the same composition and have been prepared in the same condition.
Coating rates of these samples differ while final thickness of the
samples is the same ($250nm$). We have considered copper thin films.
Copper thin films have different applications in various
technologies such as optics and laser science, because of the high
reflection power of copper in red and infrared region of spectrum.
Glass substrates were used in this survey because of smoothness. In
Tab. I, the experimental conditions of two selected thin layers are
given. AFM measurements of these samples were carried out and the
exported data was used for further calculations. The topography of
the samples was investigated using Park Scientific Instruments
(model Autoprobe CP). The images were collected in a constant force
mode and digitized into $256 \times 256 $ pixels with scanning
frequency of $0.6$ Hz. A rough surface can be described
mathematically as $h(x)$, where $h(x)$ is the surface height of a
rough surface with respect to a smooth reference surface defined by
a mean surface height and $x$ is the position vector on the
surface.

To study the effect of topography and scaling properties of the thin
films, the standard fractal analysis is used \cite{Barabasi,mahsa2}.
When long-range correlations are absent in $h(x)$, short-range
correlations may exist. In this case, it can be more suitable to study
multifractality by Markov analysis \cite{Fazeli,Friedrich}.
Fazeli at al. showed how Markov processes play a fundamental role
in probing rough surfaces and characterizing their topography. They explained
tip convolution in AFM images by non-Markovian properties in the AFM images reported.
Our results show that by increasing the coating rates, correlation
length (grain sizes) and Markov length are decreased, roughness
exponent is decreased and our surfaces get more multifractal which
means that our probability density functions get more non-Gaussian.

The paper is organized as follows. Standard fractal method and
Markov analysis are described in Section II. Data description and
analysis based on these methods for two selected copper thin films
are given in Section III. In Section IV, we present our conclusion.

\begin{figure}[t]
\includegraphics[width=6cm,height=6cm,angle=0]{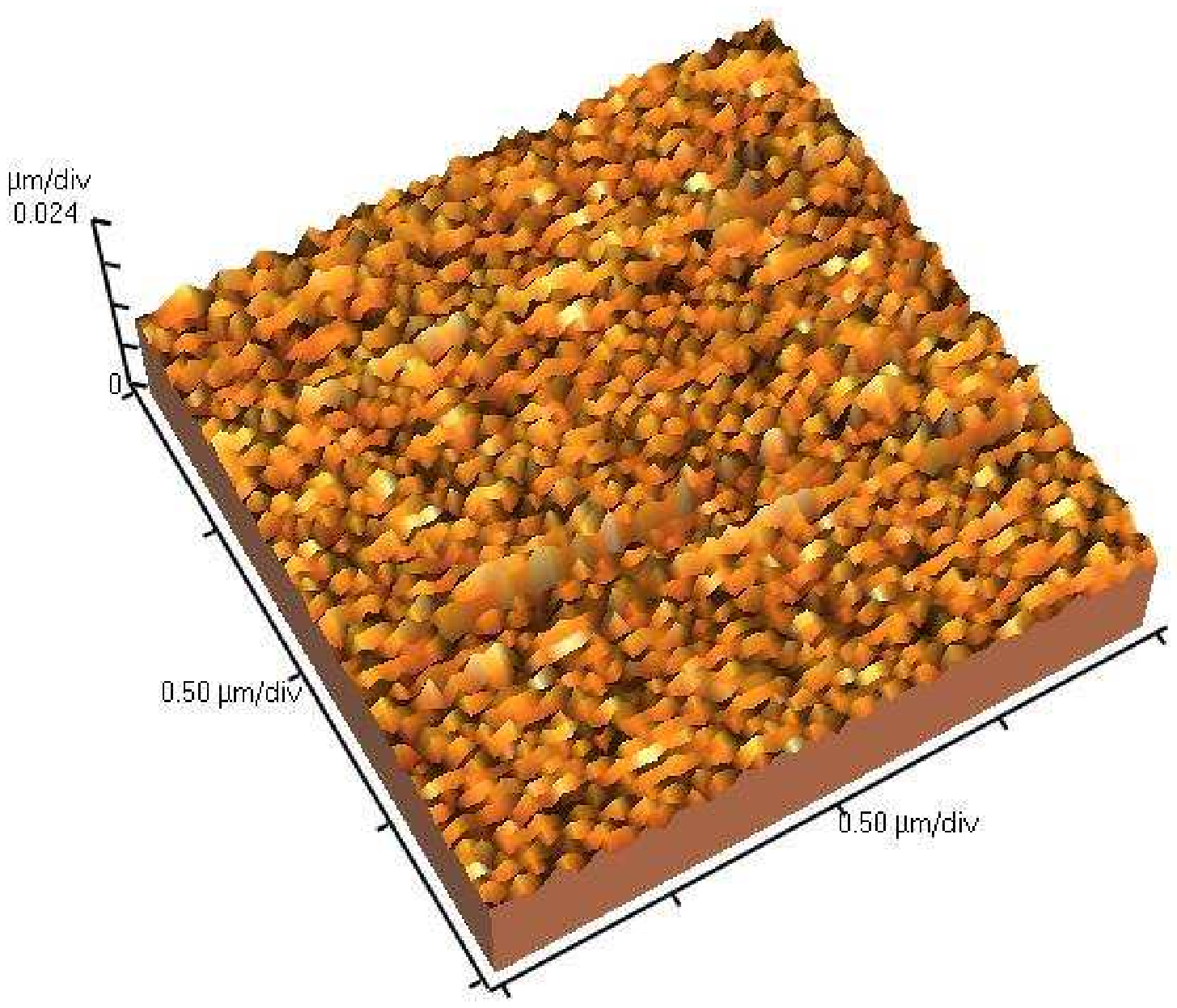}
\includegraphics[width=6cm,height=6cm,angle=0]{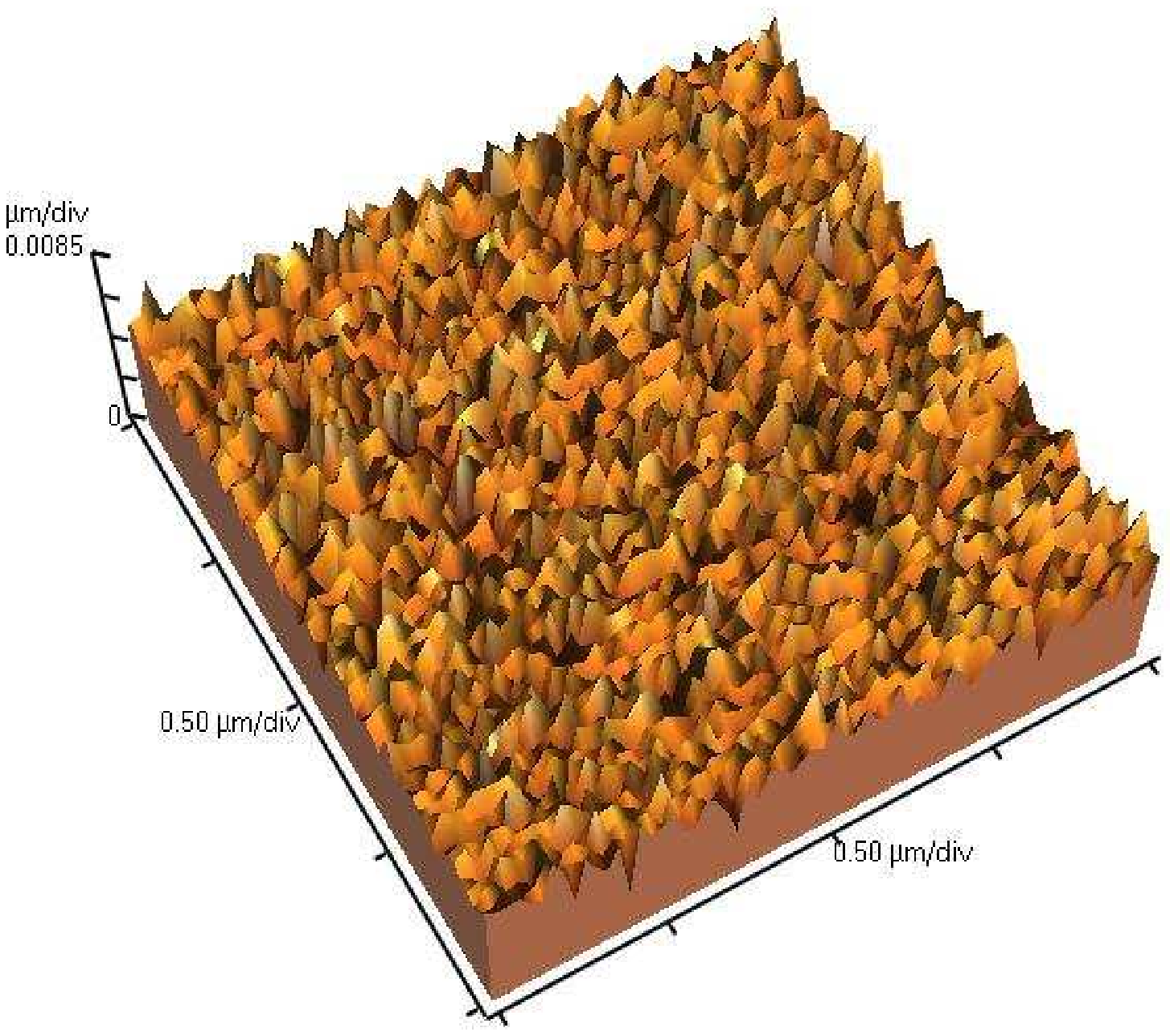}
\caption{(Color online) Atomic force microscopy (AFM) surface images
of two samples: (a) sample $1$ represents a surface with roughness
$5.2 nm$ and (b) sample $2$ represents a surface with roughness $3.6
nm$.}\label{fig2}
\end{figure}

\section{Statistical quantities}

\subsection{standard fractal analysis}
For a sample of size $L$, roughness is defined by $w(L) =(\langle
(h-\overline{h})^2\rangle)^{1/2}$ \cite{Barabasi}, where
$\langle\cdots\rangle$ denotes an spatial averaging
samples, respectively. For simplicity, without losing the
generality of the subject, we can assume that the mean height of the
surface is zero, $\overline{h}=0$.

Roughness is one of the scaling properties of the surface. Roughness
scales by size of the system as $w(L)\sim L^{\alpha}$, where
$\alpha$ is the roughness exponent. The common procedure to measure
the roughness exponent of a rough surface is based on a second
moment of height difference function defined as
$S^2(l)=\langle|h(x+l)-h(x)|^2\rangle$. This is equivalent to the
statistics based on the height-height correlation function
$C(l)=<h(x+l)h(x)>$ for stationary surfaces, i.e.
$S^2(l)=2(w^2-C(l))$. The second order height difference function
$S^2(l)$, scales with $l$ as $ l^{\xi_2}$ where $\alpha=\xi_2 /2$.

Assuming statistical translational invariance, different moments of
the height difference functions $S^{q}(l)=<|h(x+l)-h(x)|^{q}>$, (qth
moment of the increment of the rough surface height fluctuation
$h(x)$) will depend only on the space difference of heights $l$, and
has a power-law behavior if the process has the scaling property
\begin{eqnarray}
S^{q}(l)=<|h(x+l)-h(x)|^{q}> \propto
S^{q}(L_{0})\left({l\over\L_{0}}\right)^{\xi(q)},
\end{eqnarray}
where $L_{0}$ is the fixed largest length scale of the system, $<
\cdot \cdot \cdot>$ denotes a statistical average (for
non-overlapping increments of length $l$), $q$ is the order of the
moment (we take here $q > 0$), and $\xi_{q}$ is the exponent of the
height difference function. The main property of a multifractal
process is its characterization by a non-linear $\xi_{q}$ function
of $q$. Monofractals are the generic result of the linear behavior.
For instance, for Brownian motion (Bm) $\xi_{q} = q/2$, and for
fractional Brownian motion (fBm) $\xi_{q} \propto q$.

\subsection{Markov analysis}

For better understanding the multifractality features, it can be
useful to investigate the multifractality by Markov analysis. When
long-range correlations are absent in $h(x)$, short-range
correlations may exist. In this case, it can be more suitable to
study multifractality by this approach. As a measure of surface
roughness and to check the multifractal nature of rough surfaces, we
check the Markovian nature of the increments which is defined by
$h_{l}(x) = h(x + l) - h(x)$ depending on the length scale $l$.

First, we check whether $h_{l}(x)$ represents a Markov process
\cite{Friedrich,Waechter,Jafari,Peinke,Shayeganfar,Shayaganfar2}. If so, we estimate the
Markov Length scale (ML); the minimum length interval over which
$h_{l}(x)$ can be represented by a Markov process. For a Markov
process, knowledge of $P(h_{2}, l_{2}|h_{1},l_{1})$ is sufficient
for generating the entire statistics of $h_{l}(x)$, encoded in the
n-point PDF that satisfies a master equation that, in turn, is
reformulated by a Kramers-Moyal (KM) expansion,
\begin{eqnarray}
{\partial\over\partial l}p(h,l|h_0,l_0) = \sum_{k=1}^\infty
\left(-{\partial\over\partial h}\right)^k
[D_k(h,l)p(h,l|h_0,l_0)].\label{moyal2}
\end{eqnarray}
In order to obtain the drift ($D_1$) and diffusion coefficient ($D_2$)) for Eq. (2)
we proceed in a well defined way like it was already expressed by Kolmogorov \cite{Friedrich1,Friedrich2,Waechter}.
The conditional moments $D_k(h,l)$ for finite step sizes
$\Delta x$ are directly estimated from the data via moments of the conditional probabilities.
\begin{eqnarray}
D_k(h,l) = \lim_{\Delta x \rightarrow 0}  {l\over k!\Delta x}
\int_{-\infty}^\infty (h^\prime-h)^k p(h^\prime,l - \Delta x|h,l)
d{h^\prime}.\label{moyal3}
\end{eqnarray}

For a general stochastic process, all the KM coefficients may be
nonzero. However, provided that $D_4(h,l)$ vanishes or is small
compared to the first two coefficients \cite{Risken}, truncation of
the KM expansion after the second term is meaningful in the
statistical sense. For our samples, $D_4(h,l)$ is two orders of
magnitude less than $D_2(h,l)$. Thus, we truncate the KM expansion
after the second term, reducing it to a Fokker-Planck (FP) equation.
According to the Ito calculus \cite{Risken}, the FP equation is
equivalent to a Langevin equation,
\begin{eqnarray}
{\partial\over\partial l}h(l) = D_1(h,l) +
\sqrt{D_2(h,l)}f(l),\label{moyal6}
\end{eqnarray}
where $f(l)$ is a random force with zero mean and Gaussian
statistics, $\delta$ -correlated in h, i.e., $<f(l)f(l')> =
2\delta(l- l')$. $D_{1}$ is drift and $D_{2}$ is diffusion
coefficients. To better clarify these variables ($D_{1}$ and
$D_{2}$), we can pay attention to the Langvin equation (Eq. 4).
Here, $D_{1}$ indicates an average height difference by walking on the
surface ($\delta h=D_{1} \delta l$) and $D_{2}$ plays the role of the
variance (uncertainty) of this height hanging ($f(l) D_{2} \delta l =\delta
h$). To establish a relation between Markov approach and fractal
distributions we start with the Fokker-Planck equation for PDF of
the height increments
\begin{eqnarray}
\Delta x {\partial\over\partial \Delta x} p(h,l) =
[-{\partial\over\partial h} D_1(h,l) + {\partial^{2}\over\partial
h^{2}} D_2(h,l) ] p(h,l), \label{moyal7}
\end{eqnarray}
with the corresponding Langevin equation given by Eq.
(\ref{moyal6}). It can be shown that for any series with the type of
the correlations that is described by the self-affine distributions,
the drift and diffusion coefficients of the increment series are
given by \cite{Friedrich1,Friedrich2}
\begin{eqnarray}
 D_1(h,l) \simeq -H h\nonumber\\
 D_2(h,l) \simeq b h^2\label{moyal9}
\end{eqnarray}
where $H$ is the Hurst exponent which refers to the
first moment exponent and $b$ indicates the strength of the
multi-fractality. It means if $b=0$ in this case we find the samples
mono-fractal.
 Thus, using Eqs. (\ref{moyal7}), we
obtain the evolution of the moments of height difference function,
$S_q(l)\equiv \langle|Delta h(l) |^{q}\rangle=
\langle|h(x+l)-h(x)|^{q}\rangle$, as follows
\begin{eqnarray}
-l{\partial\over\partial l}\langle|\Delta h(l)|^q\rangle =
q\langle|\Delta
h(l)|^{q-1} D_1(\Delta h,l)\rangle\nonumber\\
+q(q-1)\langle|\Delta h(l)|^{q-2}D_2(\Delta
h,l)\rangle\label{moyal10}
\end{eqnarray}

Then, by substituting Eq. (6) in Eq. (7) the scaling behavior of the
moments of the height difference function are written as
$S_{q}(l)\sim(l)^{\xi_{q}}$ then,
\begin{eqnarray}
\xi_q = Hq - bq(q-1),\label{kisi4}
\end{eqnarray}
which establishes a direct link between the scaling exponents
$\xi_{q}$ and the results obtained for the drift and diffusion
coefficients.
\begin{figure}[t]
\includegraphics[width=16cm,height=13cm,angle=0]{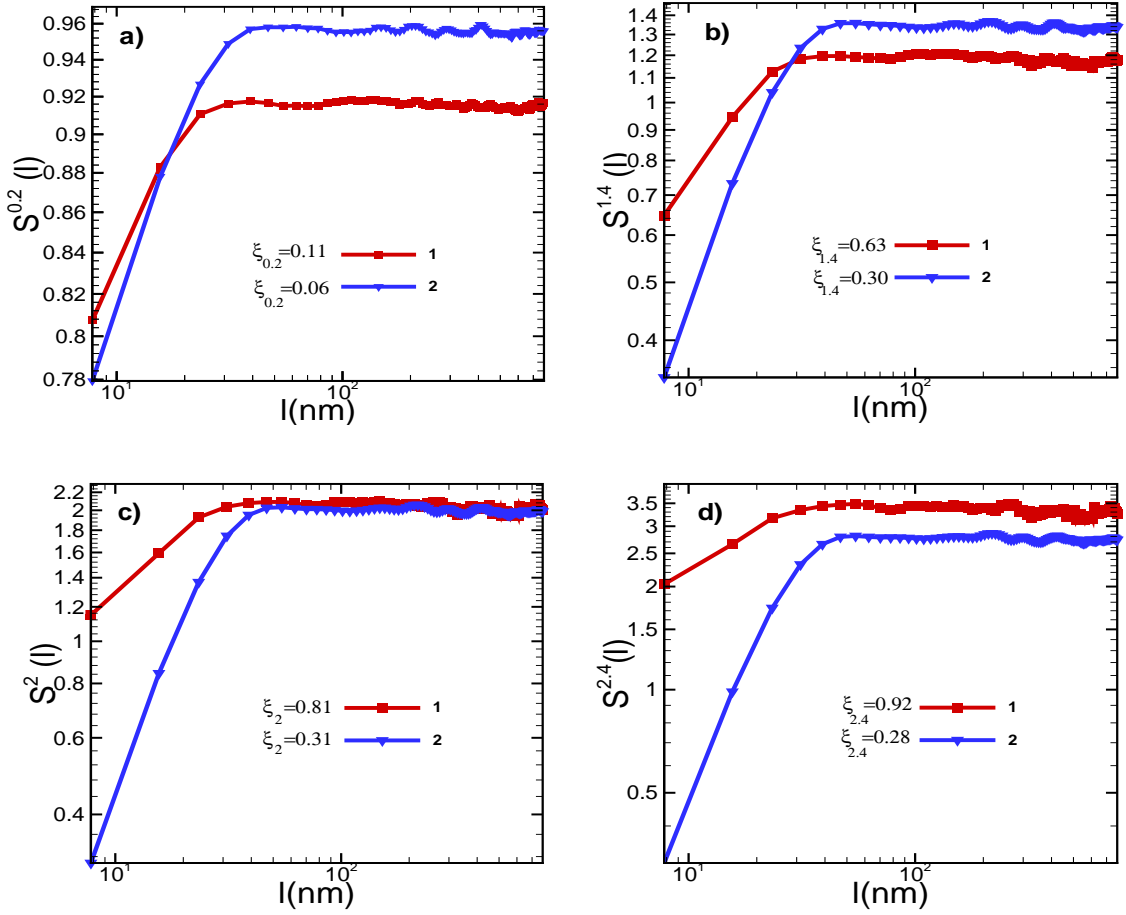}
\caption{(Color online) Height difference functions of two samples
(1 ($w=5.2nm$) and 2 ($w=3.6nm$)) for different moments q's ((a)
$q=0.2$, (b) $q=1.4$, (c) $q=2$ and (d) $q=2.4$), versus
distance.}\label{fig3}
\end{figure}
\begin{figure}[t]
\includegraphics[width=10cm,height=7cm,angle=0]{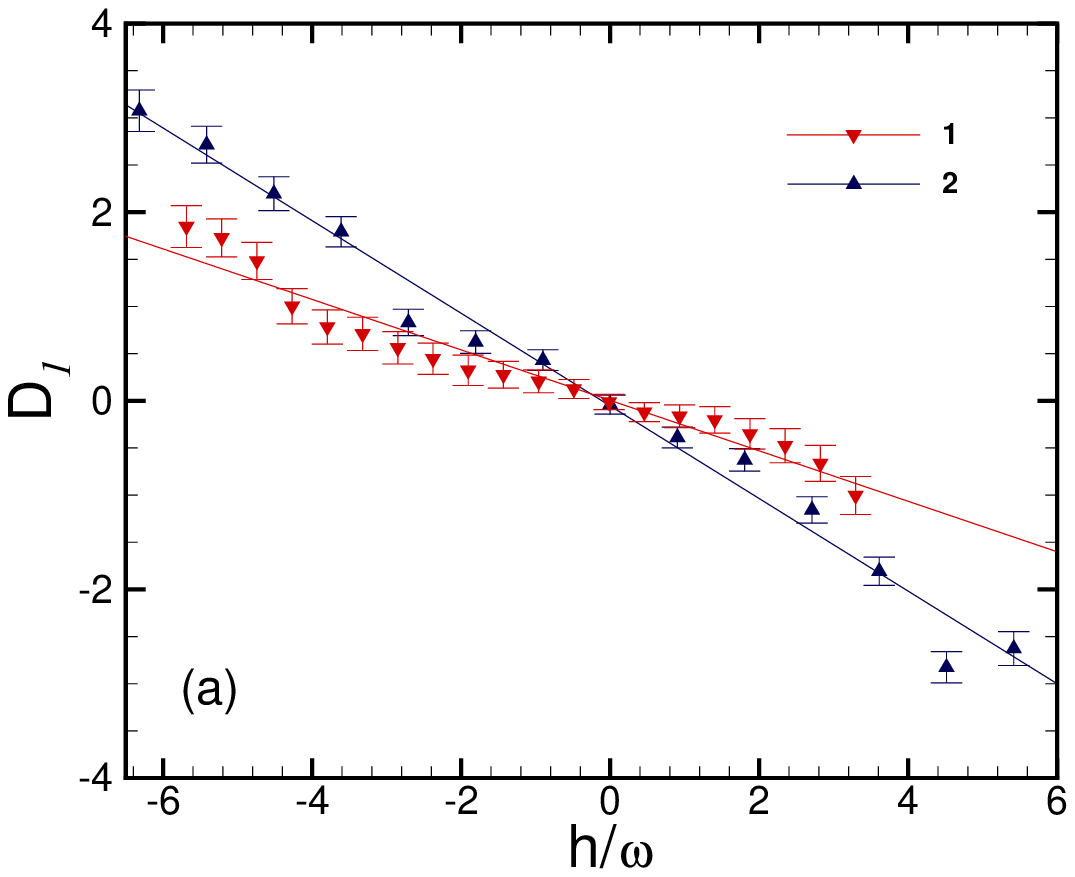}
\includegraphics[width=10cm,height=7cm,angle=0]{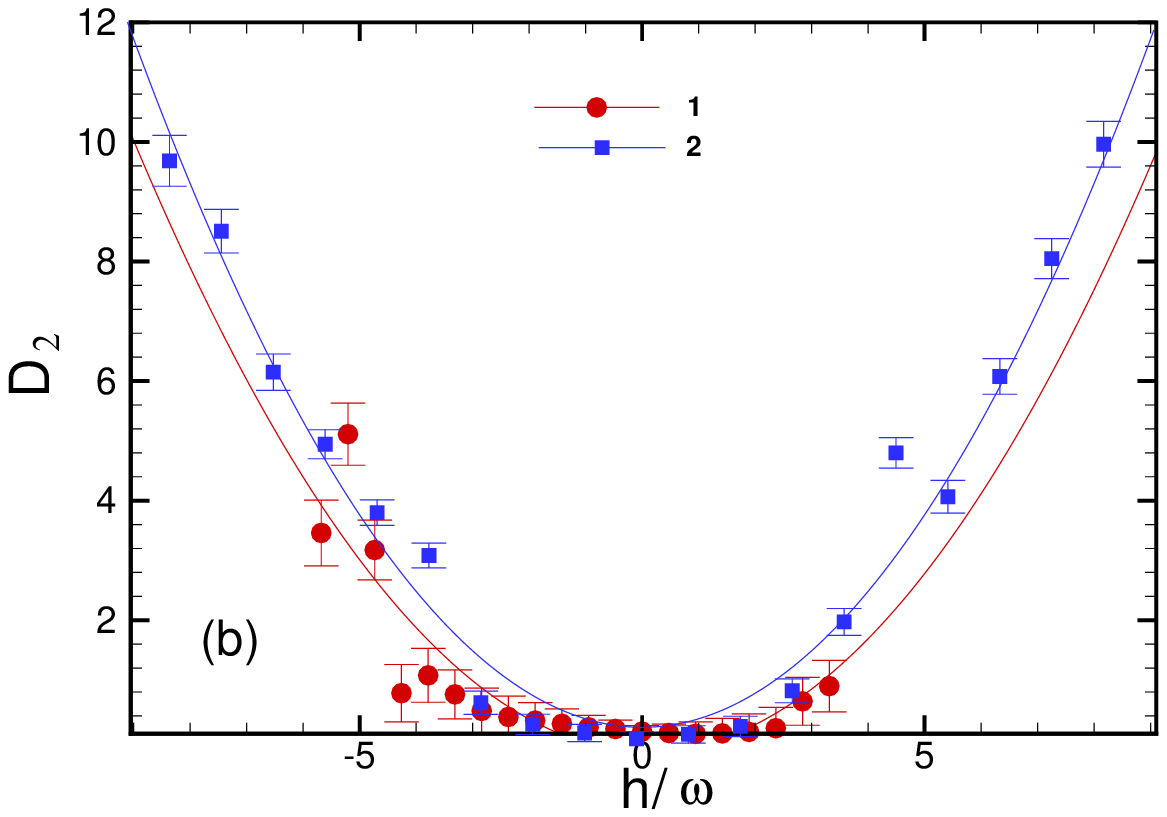}
\caption{(Color online) Drift and diffusion coefficients of the two
samples (1 ($w=5.2nm$) and 2 ($w=3.6nm$)) versus height which is
normalized by roughness ($w$). }\label{fig6}
\end{figure}
\begin{figure}[t]
\includegraphics[width=10cm,height=9cm,angle=0]{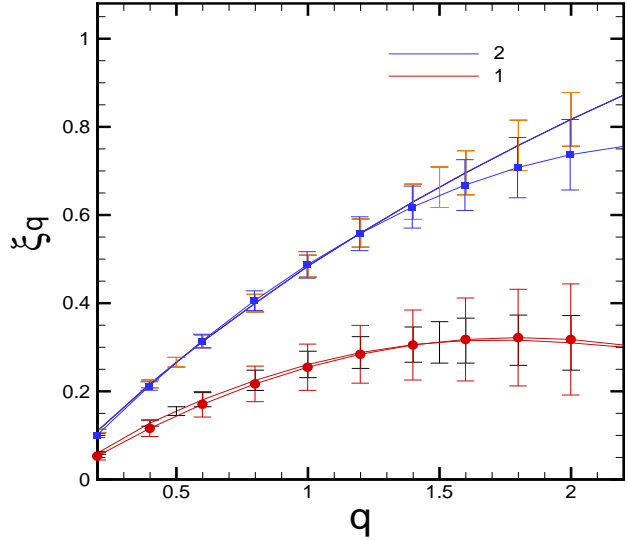}
\caption{(Color online) Exponents of the qth moments of the height
difference function of the two samples (1 ($w=5.2nm$) and 2
($w=3.6nm$)) vs q. The errors from markovian methods are shown with
the same color as the curves but the errors from the direct method
are shown with different colors (black (sample 1) and orange
(sample 2).}\label{fig7}
\end{figure}

\section{Discussion and Results}

In order to check the effect of coating rate on statistical
properties of surfaces we have used samples of the same material
(copper thin films) and equal thickness ($250 nm$) which have been
prepared in the same condition except their coating rates. We have
chosen two of these samples ($1$ and $2$) typically for drawing the
plots. In table I, experimental conditions of these selected thin
layers are given. AFM method was used for obtaining microstructural
data from the surfaces (Figs. 2a and 2b). Scaling properties were
obtained using AFM images. In order to better explain the scaling
properties of these surfaces, two points should be noted.

(a) Roughness is not an intrinsic property of the surface. Indeed,
it depends on the scale of observation. In other words, when we
observe a surface from various scales different roughness could be
obtained.

(b) All information about the roughness is not included in the
second moment of the height difference function \cite{mahsa2}.
They used the concept of the higher moments to obtain further insights on
roughness. They showed how these concepts could explain a rough
surface for the purpose of suitable application and
how experimental parameters can affect its properties.

Roughness of the two selected samples is different because of
different coating rates. Normally, roughness is defined by root mean
square of the height fluctuation (rms), however, roughness is not
only determined from the second moment of the height fluctuation
difference but also the other moments play role in the determination
of the roughness which we call "generalized roughness". Fig.
\ref{fig3} (a-d) typically presents qth moment ($0.2, 1.4, 2$ and
$2.4$) of the samples' height fluctuation versus distance. Fig.
\ref{fig3} (c) presents second moment of the samples' height
difference function. The slope of each curve yields the roughness
exponent of the corresponding surface. The scale of saturation limit
in this curve is the correlation length. At scales larger than the
correlation length the second moment saturates to a height $2w^2$.
By increasing the order of moment, $q$, we see that the height of
saturation of the moment is increased which is an indication of the
generalized roughness. As it can be seen from the figure, after a
particularhubi moment ($q>2$) the height difference function of the
first sample lies above the second one which represents that the
concept of roughness that is obtained from higher moments are larger
for the sample $1$. In other words, higher moments focus on large
height differences that occur in large scales. Thus, in large
scales, sample $1$ is rougher than the second sample. It should be
noted that the reported roughness in Tab. I presents the roughness
for the largest scale which is the size of the system.

When we have fractal systems, their $\xi_{q}$ is a linear function
of $q$ and this means that they are the same in their scaling
behavior but for systems which are multifractal this behavior
changes and the behavior of $\xi_{q}$ deviates from the simple
linear relation with respect to $q$. As can be seen in Fig.
\ref{fig3}, the scaling length, which is used to find the exponents
$\xi_{q}$, is small. Thus, in order to find the behavior of the
exponents of the height difference function moments we have used the
Markov analysis too. In addition, this method helps us to better
understand the multifractality of the surfaces under study. To use
this approach we estimate the Markov Length scale (ML), the minimum
length interval over which $h_{l}(x)$ can be represented by a Markov
Process. To estimate the Markov length we have used Chapman–Kolmogorov test. More
detailed discussions of this test can be found in the appendix of \cite{Fazeli}.
The drift and diffusion coefficients, $D_{1}$ and $D_{2}$,
were estimated directly from the data. They are well-represented by
the approximates,
\begin{equation}
D_{1}(h,l)=\left\{
  \begin{array}{ll}
    -0.51 \pm 0.01 h, & \hbox{sample 1;} \\
   -0.26 \pm 0.03 h, & \hbox{sample 2.}
  \end{array}
\right.
\end{equation}
and
\begin{equation}
D_{2}(h,l)=\left\{
  \begin{array}{ll}
    0.13\pm 0.01 h^{2}, & \hbox{sample 1;} \\
     0.10\pm 0.01 h^{2}, & \hbox{sample 2.}
  \end{array}
\right.
\end{equation}
Comparing Markov length scale and correlation length, ML is
obtained directly from the joint PDF but the correlation length is
obtained from height-height correlation function, which is the
distance at which the correlation function falls to $1/e$ of its
initial value. The Markov length scale of samples $1$ and $2$ are
$1$ pixel ($7.8 nm$) and $2$ pixel ($15.6 nm$), respectively.
Correlation length scale of samples $1$\ and $2$ are $30nm$ and
$47nm$, respectively. We summarized these quantities in Tab. I. For
sample $2$, since its Markov length is other than one, we have
considered the size effect for diffusion coefficient. The calculated
diffusion coefficient for this sample is the one with the correction
term. Non-negligible corrections have to be employed in order to get
reliable estimates of diffusion coefficient for finite Markov
length, $\Delta$ \cite{PRLkantz}.
\begin{equation}
D_{2}(h,l,\Delta)=\frac{D_{2}(h,l)-(\Delta
D_{1}(h,l))^{2}}{2\Delta(1+\Delta {D^{'}_{1}}(h,l))}.
\end{equation}

By obtaining the drift and diffusion coefficients (Fig. \ref{fig6}),
the $\xi_{q}$ behavior from both standard fractal and Markov
analysis has been calculated. Fig. \ref{fig7} shows the results from
both methods. The curves with error-bars are the ones calculated
from the Markov analysis and the other two curves are from standard
method. The equation for the $\xi_{q}$ for the two samples from the
Markov analysis are $\xi(q)=(0.26 \pm 0.03 )q - 0.10\pm 0.01 q(q-1)$
for sample $1$ and $\xi(q)=(0.51 \pm 0.01) q - 0.13\pm 0.01 q(q-1)$
for sample$2$.

As can be seen in Fig. \ref{fig7}, $\xi_{q}$ of both of our samples
deviate from lines and the one which has larger root mean square
(sample $1$) deviates more. Higher moments have more information
from the larger fluctuations, which in correlated systems larger
fluctuations appear in the larger scales. Here, $\xi_{q}$ of our
samples shows how large scales play more effective role in the
height fluctuation of sample $1$. This means that large moments of
the rougher sample grow more with respect to the other sample. In
other words, as can be seen in Fig. \ref{fig3}, in frame (a) the
$q=0.2$ moment of height difference function of sample $1$ is below
sample $2$ (after saturation length), but by increasing $q$ this
behavior changes. Lower moments of sample $2$ are higher than sample
$1$ and in the higher moments sample $1$ is higher. Lower moments
describe the small height fluctuations in the surface.

To summarize, by increasing the coating rates, correlation length
(grain sizes) and Markov length are decreased and roughness exponent
is decreased and our surfaces become more multifractal.

\begin{table}
\caption{\label{Tb1} Experimental conditions and properties of the
two selected copper thin layers: coating rate, coating time,
thickness ($\tau$), Roughness ($w$), pixel's size, correlation
length ($l_{cor}$), roughness exponent ($ \alpha $) and Markov
length scale.}
\medskip
\begin{tabular}{|c|c|c|c|c|c|c|c|c|c|}
  \hline
  Sample & Coating rate ($\frac{nm}{sec}$)&$ t(sec)$& $\tau(nm)$&$w(nm)$ & Pixel's size ($nm$) & $l_{cor}$ ($nm$)  & $\alpha$ & ML ($nm$) \\\hline
  $1$    & 1.9 $\pm $ 0.04& 128 $\pm$ 1& 250 $\pm$ 5& 5.2  $\pm $ 0.2&7.8& 31 $\pm$ 7.8&0.20 $\pm $ 0.01& 7.8 \\\hline
  $2$    & 1.7 $\pm $ 0.04& 145 $\pm$ 1& 250 $\pm $ 5& 3.6 $\pm $ 0.2&7.8&47  $\pm$ 7.8&0.40 $\pm $ 0.01& 15.6 \\\hline
  \end{tabular}
\end{table}

\section{Conclusion}

Different conditions in particle deposition leads to the change in
the topography of the surfaces. We have shown that by changing the
coating rate properties of the surface differ. The average energy of
the depositing particles is increased by raising the coating rate.
In this case, the particles penetrate through the surface and this
affects the roughness and multifractality of the surface.


\end{document}